\documentstyle[12pt]{article}
\begin{document}
\title{A BRIEF NOTE ON FLUCTUATIONS AND INTERACTIONS}
\author{B.G. Sidharth$^*$\\
B.M. Birla Science Centre, Hyderabad 500 063 (India)}
\date{}
\maketitle
\footnotetext{$^*$E-mail:birlasc@hd1.vsnl.net.in}
\begin{abstract}
In this brief note we re-emphasize the fact that an underpinning of fluctuations
characterizes the fundamental interactions in the light of El Naschie's
recent work.
\end{abstract}
In a recent paper El Naschie has introduced the concept of a fluction
\cite{r1}, a result of geometric fluctuation which could lead towards a
unification of fundamental forces. It is pointed out in this brief note that recent work by the
author does indeed emphasize the underpinning of fluctuations for fundamental
interactions.\\
In this recent work\cite{r2,r3,r4,r5,r6,r7}, it
was pointed out firstly that the fluctuation of the electromagnetic field
(or the Zero Point Field) leads to\cite{r8},
\begin{equation}
\Delta B \sim \sqrt{\hbar c}/L^2,\label{e1}
\end{equation}
where $L$ is the spatial extent. It was pointed out that if $L \sim$ Compton wavelength
of a typical elementary particle then from (\ref{e1}) we recover the mass
and energy of this particle. In other words at the Compton wavelength the
elementary particle "condenses" out of the background Zero Point Field.
Similarly a fluctuation in the metric leads to (Cf.refs.\cite{r1,r2}),
\begin{equation}
\Delta \Gamma \sim \frac{\Delta g}{L} \sim l_P/L^2\label{e2}
\end{equation}
where $l_P \sim 10^{-33}cms \sim$ Planck scale. Unlike in equation (\ref{e1}),
if $L$ in (\ref{e2}) is taken to be $\sim l_P$ then from (\ref{e2}) we
get the gravitational interaction.\\
That fluctuations tie up equations (\ref{e1}) and (\ref{e2}) can be seen
explicitly as follows. As is known, given $N \sim 10^{80}$ elementary particles
in the universe, the fluctuation in the particle number is $\sim \sqrt{N}$ which
leads to a fluctuational electromagnetic energy which in the above scheme is
the energy of the typical elementary particle, so that we have (Cf. also\cite{r9})
\begin{equation}
\frac{e^2\sqrt{N}}{R} = mc^2\label{e3}
\end{equation}
Using in (\ref{e3}) the fact that\cite{r3,r10},
$$R = \frac{GNm}{c^2}$$
we get the well known relation
\begin{equation}
e^2 \sim Gm^2 \cdot \sqrt{N} = Gm^2 \cdot 10^{40}\label{e4}
\end{equation}
Equation (\ref{e4}) is usually interpreted as an adhoc or empirical relation
comparing the strengths of gravitational and electromagnetic forces. But once
the fluctuational underpinning has been taken into account, we have deduced (\ref{e4})
and can now see the connection
between electromagnetic and gravitational interactions. Indeed from (\ref{e4})
one can deduce that\cite{r11} at the Planck scale the electromagnetic and
gravitational forces become equal, or alternatively the Planck scale of mass $\sim 10^{-5} gms$
is a Schwarszchild black hole.\\
Indeed in the model referred to earlier, elementary particles like electrons
are Kerr-Newman type black holes giving at once both the electromagnetic and
gravitational fields including the Quantum Mechanical anomalous gyro magnetic
ratio $g = 2$\cite{r12}.\\
From this point, it was shown that the strong interactions follow at the Compton
wavelength scale itself, where the dimensionality is low (Cf.ref.\cite{r6,r4,r5}).
Infact within the same scheme, it was shown that the very puzzling characteristics
of quarks namely their fractional charge, handedness and confinement besides
the order of their massses can be deduced.\\
It is by the same argument of the fluctuation of the number of particles that
it was shown that the weak interactions can also be explained\cite{r7,r13}.
Indeed similar arguments in a different context were put forward years ago
by Hayakawa\cite{r14}.\\
Briefly if the weak force is mediated by a particle of mass $M$ and Compton
wavlength $L$ we get from the fluctuation of particle number, this time
$$g^2 \sqrt{N}L^2 \approx Mc^2 \sim 10^{-14},$$
whence the weak interaction can be characterised.\\
The conclusion is that the
spirit of El Naschie's fluction is vindicated (Cf.also ref.\cite{r15}).


\begin{thebibliography}{99}
\bibitem {r1} M.S. El Naschie, CSF, 11 (2000) 1459-1469.
\bibitem {r2} B.G. Sidharth, Ind. J. of Pure and Applied Physics, 35,
p.456ff (1997).
\bibitem {r3} B.G. Sidharth, Int.J.Mod.Phys.A, 13 (15), p.2599ff (1998).
\bibitem {r4} B.G. Sidharth, Mod. Phys. Lett. A., 14 (5), pg.387ff (1999).
\bibitem {r5} B.G. Sidharth, "Universe of Chaos and Quanta", in Chaos,
Solitons and Fractals, in press. xxx.lanl.gov.quant-ph: 9902028.
\bibitem {r6} B.G. Sidharth, in Instantaneous Action at a Distance in
Modern Physics: "Pro and Contra" , Eds., A.E. Chubykalo et. al., Nova Science
Publishing, New York, 1999.
\bibitem {r7} B.G. Sidharth, "From the Neutrino to the Edge of the Universe",
to appear in Chaos Solitons and Fractals.
\bibitem {r8} C.W. Misner, K.S. Thorne and J.A. Wheeler, "Gravitation",
W.H. Freeman, San Francisco (1973).
\bibitem {r9} S. Hayakawa, Suppl. of PTP, 1965, pp532-541.
\bibitem {r10} B.G. Sidharth, Int.J.Th.Phys., 37 (4) p.1307ff (1998).
\bibitem {r11} B.G. Sidharth, "The Emergence of the Planck Scale", to appear in
Chaos Solitons and Fractals".
\bibitem {r12} B.G. Sidharth, Gravitation \& Cosmology, 4 (2) (14), 1998, 158ff.
\bibitem {r13} B.G. Sidharth, "Quantum Mechanical Black Holes: Issues and
Ramifications", Proceedings of "Frontiers of Fundamental Physics", in Press.
\bibitem {r14} S. Hayakawa, PTP, (Letters to the Editor), 1965, pp.538-539.
\bibitem {r15} M.S. El Naschie, "Towards the Unification of Fundamental
Interactions....", to appear in Chaos Solitons and Fractals.
\end{thebibliography}
\end{document}